\documentclass[10pt, conference]{IEEEtran}
\IEEEoverridecommandlockouts
\usepackage[noadjust]{cite}
\usepackage{amsmath, amssymb, amsfonts, url}
\usepackage[ruled,vlined]{algorithm2e}
\usepackage{graphicx, tabularx, booktabs, subcaption}
\usepackage{textcomp}
\usepackage{xcolor}
\usepackage[acronym]{glossaries}

\def\BibTeX{{\rm B\kern-.05em{\sc i\kern-.025em b}\kern-.08em
    T\kern-.1667em\lower.7ex\hbox{E}\kern-.125emX}}

\begin{document}

\title{The DCT Neuron for Estimation and Compensation of Amplitude Distortions in OFDM Systems
\thanks{This work is part of the project SOFIA PID2023-147305OB-C32 funded by MICIU/AEI/10.13039/501100011033 and FEDER/UE.}
}

\author{\IEEEauthorblockN{Marc Martinez-Gost\IEEEauthorrefmark{1}, Ana Pérez-Neira\IEEEauthorrefmark{1}\IEEEauthorrefmark{2}\IEEEauthorrefmark{3}, 
Miguel Ángel Lagunas\IEEEauthorrefmark{2}}
\IEEEauthorblockA{
\IEEEauthorrefmark{1}Centre Tecnològic de Telecomunicacions de Catalunya, Spain\\
\IEEEauthorrefmark{2}Dept. of Signal Theory and Communications, Universitat Politècnica de Catalunya, Spain\\
\IEEEauthorrefmark{3}ICREA Acadèmia, Spain\\
\{mmartinez, aperez, malagunas\}@cttc.es
}}

\newacronym{AI}{AI}{Artificial Intelligence}
\newacronym{AirComp}{AirComp}{over-the-air computation}
\newacronym{AWGN}{AWGN}{additive white Gaussian noise}
\newacronym{CNN}{CNN}{convolutional neural network}
\newacronym{CSI}{CSI}{channel state information}
\newacronym{DA}{DA}{direct aggregation}
\newacronym{DSB}{DSB}{double sideband}
\newacronym{FL}{FEEL}{federated learning}
\newacronym{FSK}{FSK}{frequency shift keying}
\newacronym{MSE}{MSE}{mean squared error}
\newacronym{NMSE}{NMSE}{normalized mean squared error}
\newacronym{PAM}{PAM}{pulse amplitude modulation}
\newacronym{PPM}{PPM}{pulse position modulation}
\newacronym{TBMA}{TBMA}{type-based multiple access}
\newacronym{SNR}{SNR}{signal-to-noise ratio}

\maketitle
\begin{abstract}
We present a receiver-side framework for identifying amplitude distortions in frequency-selective OFDM channels. The core novelty is the use of the DCT Neuron, a compact adaptive processor based on the discrete cosine transform (DCT), to characterize the channel’s nonlinear response, leveraging its properties for highly efficient estimation. Operating directly in the time domain, the method builds an accurate signal model and tracks channel variations adaptively, achieving reliable identification with as few as two OFDM symbols. The learned nonlinear response can then be exploited for predistortion and iterative decoding, enabling low-complexity, real-time adaptive compensation of complex responses in multicarrier systems.
\end{abstract}

\section{Introduction}

Orthogonal frequency-division multiplexing (OFDM) is a cornerstone of modern wireless communications and remains a key enabler in  future standards due to its spectral efficiency and robustness to multipath propagation. However, a well-known limitation of OFDM is its high peak-to-average power ratio (PAPR), which makes the transmitted signal particularly sensitive to nonlinear distortions introduced by power amplifiers. This leads to performance degradation and reduced power efficiency.

To address this issue, a wide range of techniques have been proposed. On one hand, transmitter-side approaches rely on digital predistortion to linearize the signal before amplification \cite{PA_modeling, fuzzy}. However, implementing a predistortion loop in practice is challenging: as system bandwidths and configurations grow, the complexity and adaptivity required make it difficult to scale efficiently, limiting its effectiveness in modern large-scale wireless networks \cite{lopez_bueno_2023}. 
On the other hand, receiver-side methods often approximate the nonlinear distortion as an additive linear impairment, which requires prior knowledge of the nonlinear response \cite{cioffi_iterative, gregorio}. More recently, deep learning approaches have been introduced to jointly compensate distortions and decode the signal in an end-to-end manner \cite{dl_geoffrey, 11104481}. While promising, these methods typically require large training datasets and lack the flexibility to adapt to time-varying conditions of the nonlinear response.

In this work, we propose a receiver-side framework to identify nonlinear frequency-selective channels in an adaptive manner. The method operates directly on the received time-domain signal and is compliant with the maximum-likelihood criterion. At the core of our approach is the DCT Neuron, an expressive computational unit that models nonlinearities using the discrete cosine transform (DCT) while dynamically adapting its coefficients via a learning rule. Thanks to the favorable properties of the DCT, the system converges extremely quickly, requiring as few as 2 OFDM symbols for accurate estimation. Once the nonlinear channel is identified, this information can be leveraged for compensation through transmitter-side predistortion or receiver-side iterative detection strategies. This work extends the approach introduced in \cite{nature_takesover} to complex-valued signals and scenarios without channel feedback effects, providing a practical and efficient solution for modern multicarrier systems.



\section{System model}

Consider the baseband time-domain OFDM symbol $\{x_n\},\ n = 0, \dots, N-1$. Each sample can be expressed as $x_n = |x_n| e^{j\phi_n}$, and we assume a bounded magnitude $0\leq|x_n|\leq 1$. We further assume that a cyclic prefix (CP) is appended at the transmitter and removed at the receiver; thus, this operation is implicit and omitted for simplicity in the notation.

The signal ${x}_n$ propagates through a nonlinear communication channel. The signal first undergoes an instantaneous memoryless nonlinear transformation $f:\mathbb{R}\rightarrow\mathbb{R}$ that depends and solely affects the magnitude of the signal. This is,
\begin{equation}
    f({x}_n)=f(|{x}_n|)\text{e}^{j\phi_n},
    \label{eq:NLC}
\end{equation}
which corresponds to a traditional AM-AM distortion produced by, for instance, solid-state power amplifiers. The integration of phase AM-PM distortions left for future work. We further assume that the nonlinearity is normalized and bounded as $0\leq|f(x_n)|\leq 1$.

The signal is transmitted across a wireless communication channel with multipath, creating a frequency-selective response. We model the channel with a finite impulse response (FIR) $\mathbf{h}\in\mathbb{C}^L$, where $L$ is the number of taps. Each channel coefficient is independent and identically distributed as $h_\ell\sim\mathcal{CN}(0,1/L)$, such that the power of the channel is normalized. The received signal is
\begin{equation}
    {y}_n = \mathbf{h}^Hf({\mathbf{x}}_n)+w_n=\sum_{\ell=0}^{L-1}h_\ell^*f({x}_{n-\ell})+w_n,
    \label{eq: rx_signal}
\end{equation}
where ${\mathbf{x}}_n=[{x}_n,{x}_{n-1},\dots,{x}_{n-L}]^T\in\mathbb{C}^L$ is a vector of $L$ consecutive samples, the nonlinearity is applied element-wise and $w_n$ is additive white
Gaussian noise (AWGN) with power $\sigma^2$. According to \eqref{eq: rx_signal} we say that the receiver observes a nonlinear channel, since the nonlinear effect could be produced by the channel as well.


The goal of the receiver is to recover the transmitted signal $\{x_n\}$. Next we propose a learning-based system that allows to characterize the nonlinear channel.

\section{Nonlinear Channel Estimation}
\label{sec:estimation}

The optimal maximum-likelihood receiver must not revert the effects of $y_n$ to recover $x_n$, but to replicate the effects over $x_n$ to obtain the received signal $y_n$ \cite{nature_takesover}.  
Unlike conventional OFDM, where the receiver has access to a subset of subcarriers for channel estimation, the proposed method operates in the time domain and therefore requires the entire OFDM symbol $\{x_n\}$ to be used for estimation.

\begin{figure*}[t]
    \centering
    \includegraphics[width=\textwidth]{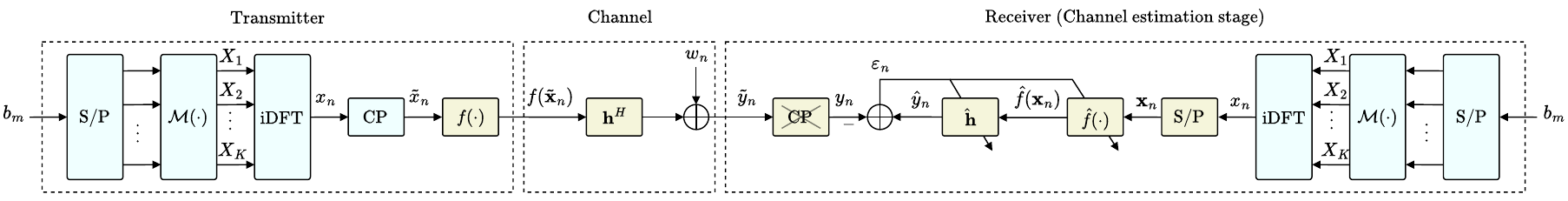}
    \caption{Time-domain estimation of nonlinear frequency-selective channels for OFDM signals.}
    \label{fig:ofdm_estimation}
\end{figure*}

Figure \ref{fig:ofdm_estimation} presents the system model as well as the receiver design to estimate the nonlinear frequency-selective channel for OFDM signals. 
Blue blocks represent the OFDM signal generation (i.e., bits $b_m$ and symbols $X_k$), while yellow blocks belong to the estimation procedure. Notice that the estimation can be extended to alternative multi-carrier systems by just exchanging the blue blocks. 

\subsection{Frequency-Selective Response}
This section focuses on the estimation of the linear channel coefficients $\mathbf{h}$. We assume that the receiver has an approximate estimation of the nonlinear function, which we denote as $\hat{f}(\cdot)\approx f(\cdot)$. 
Furthermore, we consider the channel length $L$ to be known at the receiver or long enough to be truncated afterwards.

Given a received signal $y_n$, the receiver computes $\hat{f}(\mathbf{x}_n)=[\hat{f}(x_n),\hat{f}(x_{n-1})\dots, \hat{f}(x_{n-L})]^T$, which
is fed through a linear FIR filter,
\begin{equation}
    \hat{y}_n = \hat{\mathbf{h}}^H\hat{f}(\mathbf{x}_n)=\sum_{\ell=0}^{L-1}\hat{h}_\ell^*\hat{f}(x_{n-\ell}),
\end{equation}
and $\hat{\mathbf{h}}\in\mathbb{C}^L$ is the linear channel response estimation that the receiver aims to replicate. Finally, the receiver generates the error signal that will be used to adapt the filter coefficients:
\begin{equation}
    \varepsilon_n=y_n-\hat{y}_n=y_n-\hat{\mathbf{h}}^H\hat{f}(\mathbf{x}_n),
    \label{eq:error}
\end{equation}
which corresponds to a Wiener filter and has a closed-form solution, $\hat{\mathbf{h}}=\mathbf{R}_{f}^{-1} \mathbf{r}_{fy}$, where $\mathbf{R}_{f}$ is the autocorrelation matrix of $\hat{f}(\mathbf{x}_n)$ and $\mathbf{r}_{fy}$ is the cross-correlation between the received signal $y_n$ and the input to the filter $\hat{f}(\mathbf{x}_n)$.


\subsection{Nonlinear Magnitude Distortion}
Following the same procedure as for the FIR filter, the receiver must take the reference signal $\mathbf{x}_n$ and replicate the nonlinear magnitude distortion. Specifically, we propose the DCT Neuron to model $\hat{f}(\cdot)$ as
\begin{equation}
    \hat{f}(z)=\sum_{q=1}^{Q}F_q\cos\left(\frac{\pi}{2N_\text{DCT}}(2q-1)(2z+1)\right),
    \label{eq:dct}
\end{equation}
where $F_q\in\mathbb{R}$ are the DCT coefficients, $Q$ is the total number of coefficients,  $z\in[0,N_{\text{DCT}}-1]$ and $N_{\text{DCT}}$ is the number of samples that defines the resolution of the DCT.

The objective of the adaptation is to learn the coefficients $F_q$ that minimize the error between the received signal and the replica obtained by the receiver. We refer the reader to \cite{dct_apn} for an exhaustive description of the DCT model for regression tasks and its gains in terms of convergence guarantees and sample complexity. 




Finally, equation \eqref{eq:dct} can be expressed as $\hat{f}(x_n)=\mathbf{f}^T\mathbf{c}_n \text{e}^{j\phi_n}$, where $\mathbf{f}=[F_1,F_2,\dots,F_Q]\in\mathbb{R}^Q$ contains the $Q$ DCT coefficients and $\mathbf{c}_n\in\mathbb{R}^Q$ contains the cosines evaluated at $|x_n|$. We can generalize this notation when the input is a vector $\mathbf{x}_n$ to $\hat{f}(\mathbf{x}_n)=\mathbf{f}^T\mathbf{C}_n\odot\text{e}^{j\boldsymbol{\phi}_n}$, where the $\ell$th column of $\mathbf{C}_n\in\mathbb{R}^{Q\times L}$ is $\mathbf{c}_{n-\ell}$, this is, the cosine representation for sample $x_{n-\ell}$. The vector $\text{e}^{j\boldsymbol{\phi}_n}=[\text{e}^{j\phi_n}, \text{e}^{j\phi_{n-1}},\dots, \text{e}^{j\phi_{n-L}}]\in\mathbb{C}^L$ contains the phases of the original reference signal.

Then, $\hat{f}(\mathbf{x}_n)$ is fed to the FIR channel estimate $\hat{\mathbf{h}}$, and
the instantaneous error in \eqref{eq:error} becomes
\begin{align}
    \varepsilon_n= y_n - \hat{\mathbf{h}}^H \mathbf{C}_n^T \mathbf{f}\odot\text{e}^{j\boldsymbol{\phi}_n}=y_n - \mathbf{f}^T\mathbf{C}_n\hat{\mathbf{h}}^* \odot\text{e}^{j\boldsymbol{\phi}_n},
    \label{eq:mse_total}
\end{align}
which also corresponds to a Wiener filter in terms of $\mathbf{f}$. 


To enable a simple and adaptive implementation, we minimize the error using the stochastic least mean squares (LMS) algorithm.
We define $\mathbf{u}_n=\mathbf{C}_n\hat{\mathbf{h}}^* \odot\text{e}^{j\boldsymbol{\phi}_n}$ and compute the gradient of the squared instantaneous error:
\begin{align}
    \frac{\partial||\varepsilon_n||^2}{\partial\mathbf{f}}
    =-2\Re\left(\mathbf{C}_n\hat{\mathbf{h}}^* \odot\text{e}^{j\boldsymbol{\phi}_n} \varepsilon_n^*
    \right),
\end{align}
which corresponds to the traditional LMS gradient, but projected onto the real component because the DCT coefficients are real.

To stabilize the convergence, the gradient is usually normalized by the power of the signal that goes into the LMS filter, which is $\mathbb{E}\{||\mathbf{u}_n||^2\}=
\hat{\mathbf{h}}^T
\mathbb{E}\{\mathbf{C}_n\mathbf{C}_n^T\}
\hat{\mathbf{h}}^*={QL}/{2}$. We have that $||\hat{\mathbf{h}}||^2=1$ is imposed by definition and $\mathbb{E}\{\mathbf{C}_n\mathbf{C}_n^T\}=1/2\,\mathbf{I}\in\mathbb{R}^{QL\times QL}$ because each cosine has a constant power of $1/2$, independent of the input magnitude $||x_n||$, and the cross-correlation between cosines is zero due to the orthogonality of DCT basis functions. This uncorrelation among DCT coefficients implies a uniform eigenvalue spread in all directions of the gradient, ensuring the fastest possible convergence rate for LMS \cite{dct_apn, enn}. 

Finally, the update rule used to adapt the DCT coefficients is
\begin{align}
    \mathbf{f}_{n+1} 
    =\mathbf{f}_{n}+\frac{4\alpha}{QL}
    \Re\left(\mathbf{C}_n\hat{\mathbf{h}}^* \odot\text{e}^{j\boldsymbol{\phi}_n} \varepsilon_n^*
    \right),
\end{align}
where $\alpha$ is usually referred to as the misadjustment and it is data independent.

In conclusion, the DCT not only extends the modeling capabilities of a linear filter but also integrates seamlessly with the learning dynamics of adaptive algorithms.

\subsection{General Channel Estimation Algorithm}

We therefore adopt an alternating optimization strategy, where the linear channel and the DCT coefficients are updated iteratively. Specifically, for fixed DCT coefficients, the channel is estimated in closed form, while for a fixed channel, the DCT coefficients are refined using the LMS updates. This process is repeated for $N_\text{iter}$ iterations until convergence.

\section{Signal Detection}
\label{sec:detection}

We focus on how the estimated nonlinear channel response can be used to design the detector at the receiver. Specifically, we consider the same OFDM block so that the channel’s frequency response remains constant. If this assumption does not hold, additional pilots can be inserted to estimate the multipath channel. This is feasible because the nonlinear distortion is assumed to originate from hardware effects and are slow time-variant.


In the following, we review two existing methods, which differ in whether the estimated nonlinear response is handled at the transmitter or the receiver. We then illustrate how the proposed estimation procedure can be incorporated into each approach.

\subsection{Nonlinear Predistortion at the Transmitter}
\label{sec:predistortion}

The transmitter can precompensate the signal $\{x_n\}$ and linearize the nonlinear magnitude distortion. At the receiver side, the frequency-selective channel can be compensated in the frequency domain as in conventional OFDM.

So far, the receiver has estimated $f(\cdot)$, which can be sent to the transmitter. With that, precompensation can be applied, but requires to apply the inverse $\hat{f}^{-1}(\cdot)$. We adopt a self-supervised approach in which a known input sequence $\{x_n\}$ is passed through $\hat{f}(\cdot)$ and then through a parametrized inverse model, yielding $\hat{x}_n = \hat{f}^{-1}(\hat{f}(x_n))$. The same input serves as the reference, leading to the error
\begin{equation}
    \varepsilon_n = x_n - \hat{x}_n = x_n - \mathbf{g}^T \mathbf{c}_n,
\end{equation}
where $\mathbf{g}\in\mathbb{R}^Q$ are the DCT coefficients of the inverse model. The inverse can also be learned in an adaptive fashion, which preserves the favorable convergence properties of the DCT model. While a larger number of coefficients $Q$ may be required, the procedure is noise-free and local, so performance is determined solely by the model’s approximation capacity.

\subsection{Iterative-decoding}
As proposed in \cite{cioffi_iterative}, when the receiver knows the nonlinear distortion and the channel response, it can implement a quasi-maximum likelihood detection. These algorithms assume the decomposition $f(x_n)=x_n + d_n$,
where $d_n$ encapsulates the nonlinear component of the signal and treated as noise. This linearization allows to work with OFDM in the frequency domain, as if there was no distortion. First, hard-decoding is applied over each subcarrier to obtain an estimate of the transmitted digital symbol. Then, the additive distortion $d_n$ is estimated in the time domain and used to correct the decoding in an iterative fashion.

\section{Simulation Results}
\label{sec:results}

In this section, we evaluate the proposed methods through numerical simulations. First, we assess the performance of the estimation procedure for nonlinear frequency-selective channels. Then, we investigate the performance of the proposed detectors in OFDM systems, demonstrating how the estimated models can be used to mitigate nonlinear distortions during signal detection. Without loss of generality, we consider that the estimation and detection happen within the same coherence block, so that the nonlinear channel response does not change. 

\subsection{Estimation of Nonlinear Frequency-Selective Channels}

\begin{figure}[t]
    \centering
    \begin{subfigure}[b]{0.45\columnwidth}
        \centering
        \includegraphics[width=\textwidth]{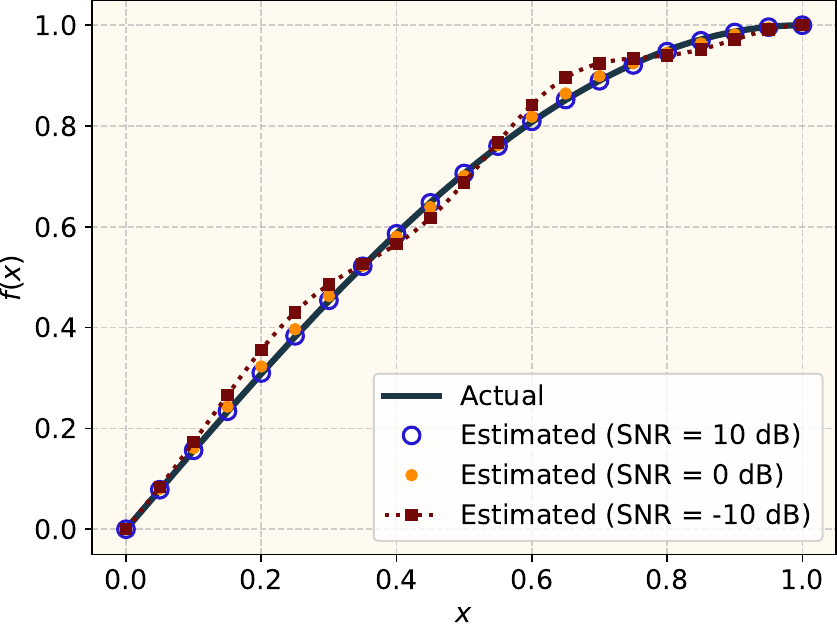}
        \caption{Soft distortion.}
        \label{fig:sine}
    \end{subfigure}
    \hfill
    \begin{subfigure}[b]{0.45\columnwidth}
        \centering
        \includegraphics[width=\textwidth]{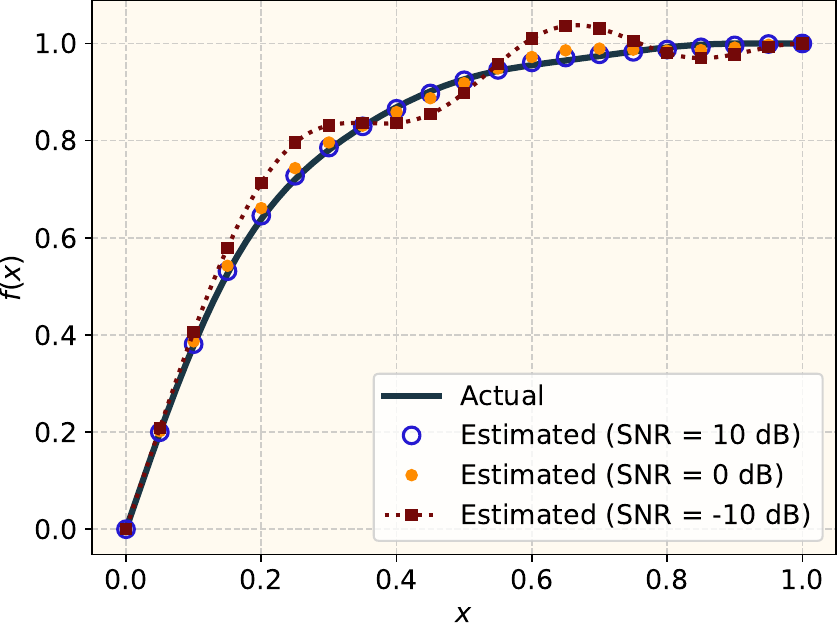}
        \caption{Hard distortion.}
        \label{fig:comp}
    \end{subfigure}
    \caption{Considered nonlinear amplitude distortions and their estimated DCT responses at different SNR levels.}
    \label{fig:amplitude_distortions}
\end{figure}

We consider a frequency-selective channel with $L=3$ taps and two types of nonlinear amplitude distortions: Figure \ref{fig:amplitude_distortions}(\subref{fig:sine}) is a soft nonlinearity, while \ref{fig:amplitude_distortions}(\subref{fig:comp}) represents a hard nonlinearity, which introduces abrupt distortion. The nonlinear channel is estimated using OFDM signals with $N=1024$ subcarriers and a 16-quadrature amplitude modulation (16-QAM) on each symbol. The adaptive algorithm is configured with $\alpha=0.01$, $N_\text{iter}=30$ and $Q=6$ coefficients.

Figure \ref{fig:amplitude_distortions} also illustrates the corresponding DCT estimates at different SNR levels. The estimation is nearly perfect down to 
$\text{SNR}=0$ dB, and even at $\text{SNR}=-10$ dB, the small estimation errors produce negligible degradation in the detector’s performance. Finally, it is noteworthy that the estimation requires only 2 OFDM symbols (i.e., $2048$ time samples) to capture the nonlinear frequency-selective channel.

\subsection{Nonlinear Compensation in OFDM}

For each of the estimations obtained in Figure \ref{fig:amplitude_distortions}, we perform symbol detection with the proposed methods described in section \ref{sec:detection}. For predistortion, the inverse function is implemented with the DCT, configured with $Q=N_\text{DCT}=512$ coefficients (i.e., all of them). The learning takes 10.000 samples with $\alpha=0.01$. The iterative-decoding algorithm is implemented with $N_\text{iter}=5$, as described in the seminal paper \cite{cioffi_iterative}.

\begin{figure}[t]
    \centering
    \begin{subfigure}[b]{0.99\columnwidth}
        \centering
        \includegraphics[width=\textwidth]{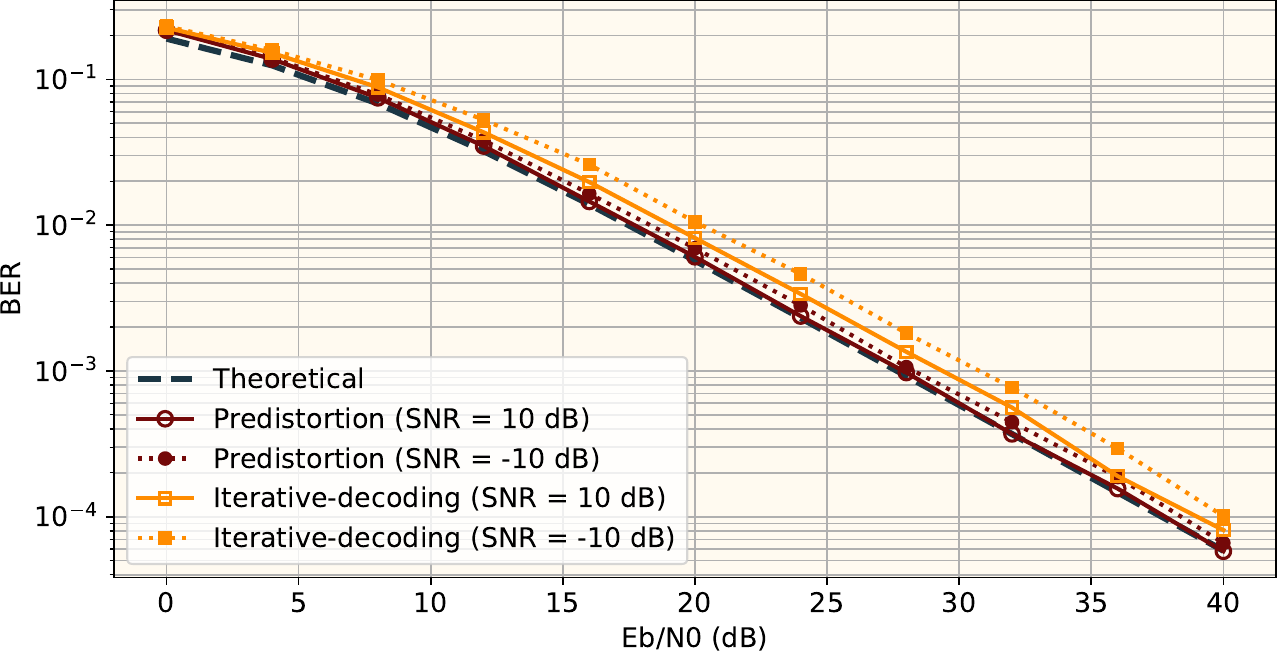}
        \caption{Nonlinear distortion of Figure \ref{fig:amplitude_distortions}(\subref{fig:sine}).}
        \label{fig:ber_sine}
        \vspace{10pt}
    \end{subfigure}
    \hfill
    \begin{subfigure}[b]{0.99\columnwidth}
        \centering
        \includegraphics[width=\textwidth]{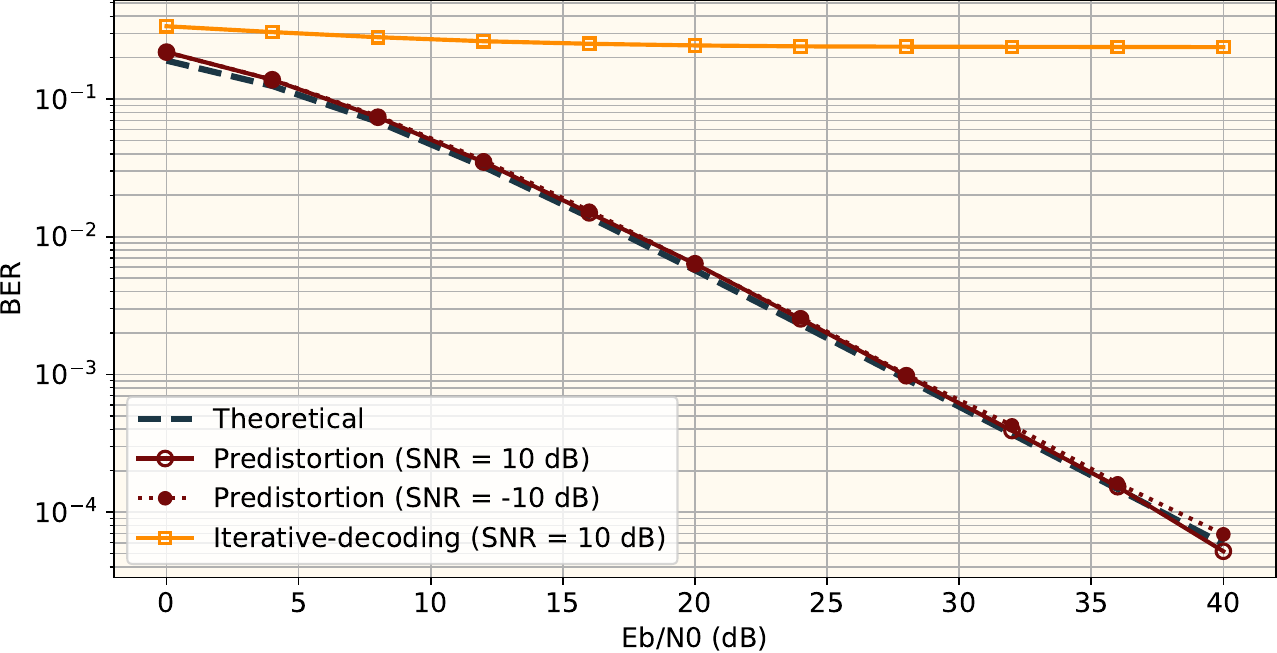}
        \caption{Nonlinear distortion of Figure \ref{fig:amplitude_distortions}(\subref{fig:comp}).}
        \label{fig:ber_comp}
    \end{subfigure}
    \caption{BER for nonlinear distortions and different decoding schemes when the estimation is performed at different SNR regimes.}
    \label{fig:ber}
    \vspace{-10pt}
\end{figure}

Figure \ref{fig:ber}(\subref{fig:ber_sine}) illustrates the raw bit error rate (BER) for the mild nonlinearity. The theoretical curve shows the BER for a linear transmission over a selective-fading channel. The predistortion technique is more robust because it completely linearizes the channel. Remarkably, even at $\text{SNR}=-10$ dB, where the nonlinearity is not perfectly estimated, the linearization is good enough to prevent errors in the decoder. On the other hand, the iterative-decoder experiences a mild degradation with respect to the lower bound and it is more sensible to the training SNR because the implicit signal model considers an equivalent noise that merges both the contributions from the channel and from the nonlinear distortion.

Figure \ref{fig:ber}(\subref{fig:ber_comp}) shows the BER for the harsh nonlinearity. As expected, the predistortion technique linearizes the response at high and low SNR regimes. Conversely, the iterative-decoding strategy does not accomplishes decoding even with infinite SNR. This result is expected because the nonlinear distortion is so large that the linearization reduces the effective SNR up to a level where most of the symbols are incorrectly decoded and the iterative procedure cannot further re-estimate the distortion.

\section{Conclusion}
In this work, we presented a receiver-side framework to identify amplitude distortions in frequency-selective OFDM channels using the DCT Neuron. By modeling all effects in the time domain, we enable accurate characterization and signal modeling, making time-domain compensation the optimal strategy. Compared to transmitter-side estimation, our approach mitigates additional radio-frequency distortions which are typically unaccounted for in predistortion schemes. 
Finally, our system highlights the importance of nonlinear identification over straightforward compensation, enabling practical, low-latency and robust real-time systems.

\bibliographystyle{IEEEbib}
\bibliography{refs}

\end{document}